\documentclass[letterpaper]{article} 
\usepackage{aaai25}  
\usepackage{times}  
\usepackage{helvet}  
\usepackage{courier}  
\usepackage[hyphens]{url}  
\usepackage{graphicx} 
\usepackage{natbib}  
\usepackage{caption} 
\usepackage{appendix}
\usepackage{blindtext}
\usepackage[bottom]{footmisc}
\usepackage{graphicx}
\usepackage{epstopdf}
\usepackage{subcaption}
\usepackage{amsmath,amsfonts,bm}
\usepackage{amssymb}
\usepackage{amsthm}
\usepackage{cases}
\usepackage{multirow}
\usepackage{array}
\usepackage{textcomp} 
\usepackage{url}
\usepackage{verbatim}
\usepackage{mathtools}
\usepackage[dvipsnames]{xcolor}
\usepackage[ruled, linesnumbered, noend]{algorithm2e}
\usepackage{tabulary}
\usepackage{booktabs}
\usepackage{makecell}

\DeclareMathOperator*{\argmax}{arg\,max}
\newtheorem{theorem}{Theorem}

\newtheorem{definition}{\textbf{Definition}}

\newcommand\numberthis{\addtocounter{equation}{1}\tag{\theequation}}

\usepackage{adjustbox}

\title{DualGFL: Federated Learning with a Dual-Level Coalition-Auction Game}

\author{
   Xiaobing Chen\textsuperscript{\rm 1},
   Xiangwei Zhou\textsuperscript{\rm 1},
   Songyang Zhang\textsuperscript{\rm 2},
   Mingxuan Sun\textsuperscript{\rm 3}
}
\affiliations {
   \textsuperscript{\rm 1}Division of Electrical and Computer Engineering, Louisiana State University\\
   \textsuperscript{\rm 2}Department of Electrical and Computer Engineering, University of Louisiana at Lafayette\\
   \textsuperscript{\rm 3}Division of Computer Science and Engineering, Louisiana State University\\
   \{xchen87, xwzhou\}@lsu.edu, songyang.zhang@louisiana.edu, msun@csc.lsu.edu
}

\begin{document}

\maketitle

\begin{abstract}
Despite some promising results in federated learning using game-theoretical methods, most existing studies mainly employ a one-level game in either a cooperative or competitive environment, failing to capture the complex dynamics among participants in practice. 
To address this issue, we propose \textbf{DualGFL}, a novel \textbf{F}ederated \textbf{L}earning framework with a \textbf{Dual}-level \textbf{G}ame in cooperative-competitive environments. 
DualGFL includes a lower-level hedonic game where clients form coalitions and an upper-level multi-attribute auction game where coalitions bid for training participation.
At the lower-level DualGFL, we introduce a new auction-aware utility function and propose a Pareto-optimal partitioning algorithm to find a Pareto-optimal partition based on clients' preference profiles.
At the upper-level DualGFL, we formulate a multi-attribute auction game with resource constraints and derive equilibrium bids to maximize coalitions' winning probabilities and profits. 
A greedy algorithm is proposed to maximize the utility of the central server.
Extensive experiments on real-world datasets demonstrate DualGFL's effectiveness in improving both server utility and client utility.
\end{abstract}

\section{Introduction}
Federated learning enables decentralized model training across edge devices without transmitting raw data to a central server, which reduces resource costs during local training and model transmission while maintaining performance.
The conventional FedAvg approach \cite{mcmahan2017communication} reduces communication overhead by scheduling a subset of clients per training round. To address data and system heterogeneity and improve efficiency or performance \cite{10589575, Jing2024FedSCPF}, existing studies have explored various techniques, such as statistical client selection \cite{chen2022optimal, 10.1145/3485730.3485930, cao2022birds, 9252927}, heuristic client selection \cite{lai2021oort, 10.1145/3495243.3517017}, and optimization-based approaches \cite{8664630, 9579038, chen2024cost}.

In practical deployments, clients and servers have distinct objectives: clients seek access to the latest global model and potential monetary payoffs \cite{YANG2023126739}. In contrast, servers aim to incentivize high-quality data contributions within budget limits \cite{9996132, 9680715}.
Robust incentive mechanisms are therefore essential, and game theory offers a promising framework by modeling strategic interactions among rational agents, improving utilities \cite{10054499, 10092911} and enhancing overall system efficiency \cite{10257241, 10279623}.

Existing game-theoretical approaches in federated learning often use one-level games, either in a purely cooperative \cite{10054499, 10257241} or competitive \cite{9372882, 10092911, 9763335, 9591335, 9057543} setting, missing complex dynamics among participants in practice.
Additionally, current methods tend to improve either client utility \cite{10054499, 10257241} or server utility \cite{9996132, YANG2023126739, 10279623, 9919199}, rarely addressing both simultaneously. This narrow focus limits their applicability in real-world, large-scale federated learning scenarios, where both cooperative and competitive dynamics are present, and the goals of the server and clients must be balanced.

Given these limitations, we pose the question: \textit{Can we design a comprehensive game-theoretical framework for practical federated learning that benefits both the server's and clients' utilities?} 
To address the question above, we propose an innovative \textbf{F}ederated \textbf{L}earning framework with a \textbf{Dual}-level \textbf{G}ame in cooperative-competitive environments, namely \textbf{DualGFL}, which meets the following criteria: 
\begin{enumerate}
    \item Clients can autonomously decide whether to participate in the training based on their utilities.
    \item The server can autonomously select participants to optimize its utility.
\end{enumerate}

To achieve this goal, we leverage the hierarchical structure in hierarchical federated learning (HFL) to develop a dual-level game-theoretical framework with both cooperative and competitive elements.
In HFL, edge servers act as intermediaries between clients and the central server, aggregating model updates from clients and transmitting the aggregated model to the central server. This hierarchical structure helps reduce communication overhead \cite{9148862}, manage system heterogeneity \cite{9054634, 9207469}, and enhance scalability \cite{9699080}.

The main contributions can be summarized as follows:
    
(1) DualGFL is the first game-theoretical federated learning framework to implement a dual-level game structure. At the lower level, a coalition formation game allows clients to autonomously assess and choose edge servers to form coalitions based on their preferences and utilities. At the upper level, a competitive multi-attribute auction game enables these coalitions to bid for participation in the training process, with the central server determining the winners to maximize its utility.

(2) Unlike existing single-level games, we introduce a new utility function for clients that considers both cooperative and competitive dynamics. This auction-aware utility function evaluates not only the payoff and cost in the lower-level game but also the expected competitiveness in the upper-level auction game.

(3) For efficient and dynamic coalition formation, we propose a Pareto-optimal partitioning (POP) algorithm to find a Pareto-optimal partition based on clients' preference profiles. POP leverages the central server for coalition formation, eliminating the need for client-to-client information exchange and achieving $O(KN^3)$ complexity.

(4) We propose a multi-attribute auction with resource constraints for the upper-level game and provide the theoretical analysis of the equilibrium bid for each coalition.

Additionally, we formulate a score maximization problem for the central server and propose a greedy algorithm with a score-to-resource ratio to solve the coalition selection.

(5) Through experiments on real-world datasets, we evaluate the performance of DualGFL on the server's and clients' utilities. Results demonstrate that DualGFL effectively improves the clients' averaged utility and significantly increases the server's utility while achieving better test accuracy than baselines with single-level games.

\section{Related Work}

Incentive mechanisms, modeled through game theory, motivate participants to engage and contribute high-quality data. Existing approaches include the following.

\textbf{Client-oriented methods} mainly involve coalition formation games where clients dynamically change coalitions to maximize utility, such as increasing profits or reducing training costs. For instance,
A coalitional federated learning scheme has been proposed where clients form coalitions via a client-to-client trust establishment mechanism \cite{10054499}.
A joint optimization problem of the user-to-edge-server association and the uplink power allocation is formulated in HFL, and a satisfaction equilibrium is found to solve the partition problem \cite{10257241}.
While effective in improving client utility, these methods suffer from high communication overhead and privacy concerns due to peer-to-peer exchanges. \textit{Our DualGFL reduces communication overhead by utilizing a central server for coalition formation, eliminating the need for extensive client-to-client exchanges}.

\textbf{Server-oriented methods} aim to improve the server's utility by incentivizing clients' data contribution and selecting high-quality clients.
Auction games are used for client selection in \cite{9372882, 10092911}.
Additional studies explore double auction mechanisms in \cite{9763335}, multi-player simultaneous games in \cite{9591335}, and private reward games in \cite{9996132} to incentivize data contributions, along with evolutionary games for cooperation and latency minimization in \cite{YANG2023126739, 10279623}.
Regarding games in HFL, mechanisms like MaxQ \cite{9992189} introduce matching games to assign clients to edge servers based on mutual preferences, and blockchain-based incentives in \cite{9919199} provide public authority and fairness.
However, these methods focus on server utility without adequately balancing client incentives. \textit{Our DualGFL addresses this problem by incorporating a dual-level game that simultaneously optimizes the utilities of both clients and the server.}

\textbf{Mixed methods} are designed to serve the utilities of both clients and the server, which is important for practical implementations. 
A Stackelberg game-based multifactor incentive mechanism for federated learning (SGMFIFL) \cite{10225303} is proposed to incentivize clients to provide more data at low costs and allow clients to change their training strategies adaptively.
Similarly, a Stackelberg game and deep reinforcement learning are used to find the optimal strategies for both clients and the server \cite{8963610}.
Existing work has shown improved results in terms of clients' and the server's utility.
However, compared with server-oriented methods, mixed methods are much less explored. 
Moreover, even in HFL, existing work only adopts single-level games in either competitive or cooperative environments, failing to model the complex interactions of participants in practice.
\textit{In our work, DualGFL integrates both cooperative and competitive dynamics through a dual-level game, enhancing the hierarchical federated learning framework by comprehensively addressing both server utility and client utility}.

\section{System Model}\label{sec:sys_models}
We introduce HFL (preliminary) followed by the cost model.

\subsection{Hierarchical Federated Learning}
In an HFL system, one central server, $K$ edge servers, and $N$ clients collaboratively train a model. The central server aggregates global models, while edge servers aggregate local models from clients. Client $i$ owns a private dataset $\mathcal{D}_i = \{\xi_j^i \mid j = 1, 2, ..., |\mathcal{D}_i|\}$ of size $|\mathcal{D}i|$, with $\mathcal{D} = \bigcup_{i \in \mathcal{N}} \mathcal{D}_i$ representing the overall data. The local loss function on dataset $\mathcal{D}_i$ is defined as
$F_i(x) \coloneqq \frac{1}{|\mathcal{D}_i|} \sum_{\xi_j^i \in \mathcal{D}_i} F_i(x; \xi_j^i),$
where $x$ represents the model parameters. HFL systems aim to find an optimal model parameter $x$ to minimize the global loss function $f(x)$:
$\min_{x} f(x) \coloneqq \sum_{i=1}^N d_i F_i(x),$
where $d_i = |\mathcal{D}_i|/|\mathcal{D}|$ represents the proportion of data held by client $i$, with $\sum_{i=1}^N d_i=1$.

In HFL, edge servers relay model updates between clients and the central server.
Assume there are $T$ global rounds needed for model convergence. 
In each round, the training process includes model broadcasting, local training, model transmission, and aggregation.
The central server randomly selects clients to participate in the $t$-th round, forming the participant set $\mathcal{M}^{(t)}$, and broadcasts the current global model parameters $x_t$ to edge servers. Edge servers relay parameters to selected clients who perform local training over $I$ iterations and send updated model parameters back to edge servers. Edge servers aggregate local models and transmit them to the central server, where global aggregation is performed:
$x_{t+1} = \sum_{i \in \mathcal{M}^{(t)}}d_iy_{t, I}^i,$
where $y_{t, I}^i$ are the updated model parameters from client $i$ after $I$ local iterations.

\subsection{Cost Model}
Model training in HFL involves both computation and communication costs for clients. The computation cost is associated with local data processing, while the communication cost involves transmitting model updates to the edge server.
Formally, we define the computation cost as
$c_{i, cp} = \kappa a_i f_i^2,$
where $\kappa$ is a coefficient based on the CPU architecture, $a_i$ is the number of CPU cycles, and $f_i$ is the CPU clock frequency \cite{8234573}.

The communication cost depends on the coalition $S$ that client $i$ is in. 
In this work, we consider a wireless environment as an exemplary application with the achievable uplink transmission rate
$r_i(S) = B\log_2(1+p_ih_i/N_0),$
where $B$ denotes the bandwidth, $p_i$ and $h_i$ are the transmission power and channel gain between client $i$ and its edge server, and $N_0$ denotes the noise power spectral density \cite{10092911}. Hence, the communication cost is
$c_{i, cm}(S) = \frac{|x|}{r_i(S)},$
where $|x|$ denotes the size of model updates.
Therefore, the total cost of client $i$
is given by
\begin{equation}\label{eq:client_cost}
    C_i(S, \theta_i) = c_{i, cp} + \theta_i c_{i, cm}(S),
\end{equation}
where $\theta_i$ denotes the cost factor of client $i$.
Note that although we define $C_i$ in the wireless scenario, our proposed DualGFL can be generalized to any networking system by customizing $C_i$ for specific applications.

\section{Dual-Level Game Federated Learning}
\subsection{Architecture}
\begin{figure}
    \includegraphics[width=\linewidth]{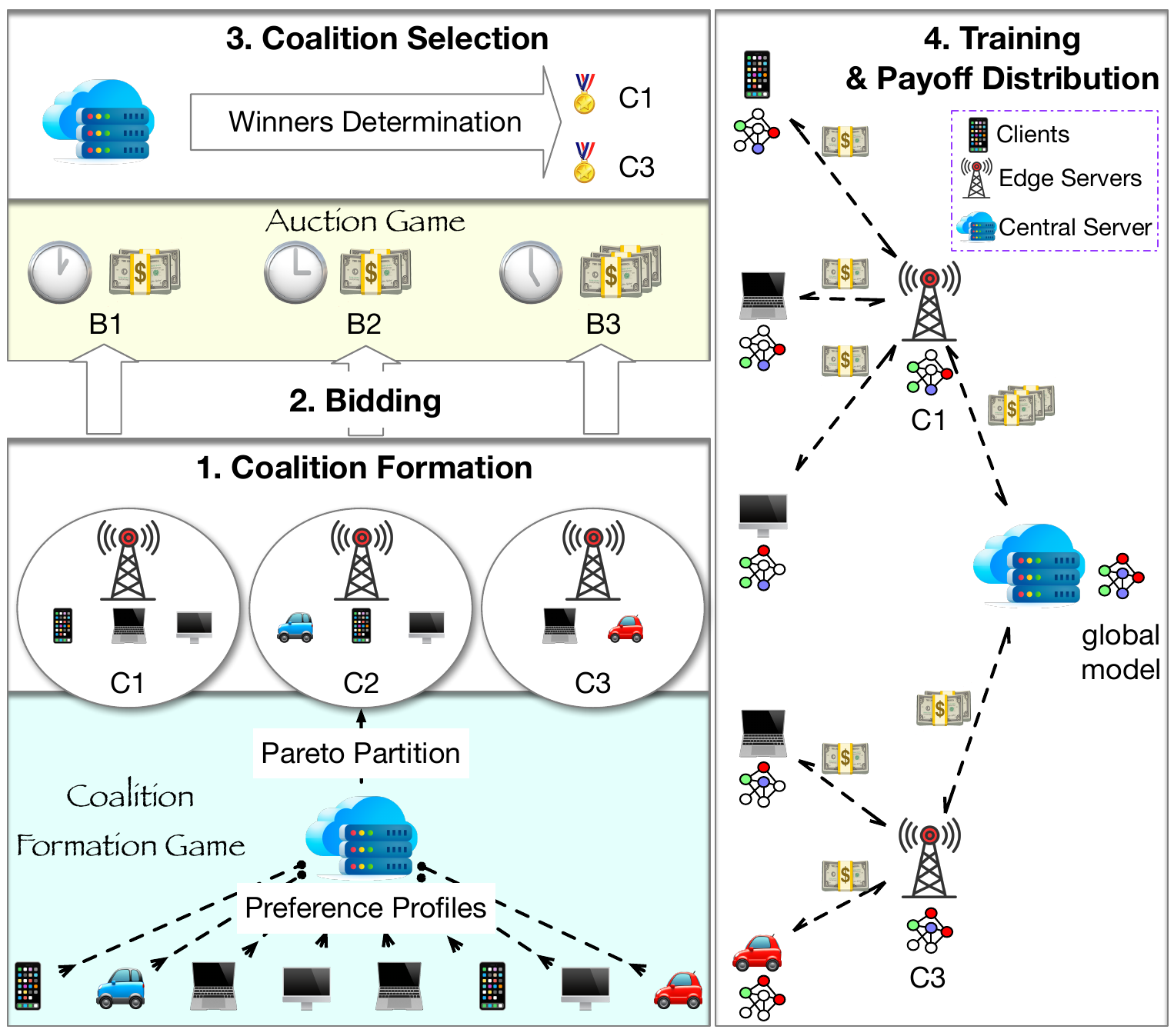}
    \captionsetup{justification=justified, singlelinecheck=false}
    \caption{DualGFL architecture.}
    \vspace{-0.25cm}
    \label{fig:gamehfl_arch}
\end{figure}

DualGFL consists of three hierarchies: clients, edge servers, and the central server, and involves two-level games: a lower-level coalition formation game and an upper-level auction game. DualGFL includes four main components: coalition formation, bidding, coalition selection, and federated training and payoff distribution, as shown in Figure \ref{fig:gamehfl_arch}.
In each training round, the following steps are executed.
(1) \textbf{Coalition Formation}: Clients calculate utilities of potential coalitions and form preference profiles.
These profiles are submitted to the central server who finds a \textit{Pareto-optimal} \cite{AZIZ2013562} partition that cannot be improved to make one client better off without making at least one client worse off.
Each coalition contains one edge server and several clients. 
(2) \textbf{Bidding}: Edge servers place bids in the upper-level auction game to participate in the training.
(3) \textbf{Coalition Selection}: The central server selects coalitions based on received bids. The winning bids are determined by the predefined rules, and every client publicly knows these rules in advance.
(4) \textbf{Federated Training and Payoff Distribution}: 
Selected coalitions receive the global model, perform local training, and send aggregated updates to the central server. The central server then distributes payoffs to coalitions, which are further distributed to individual members by the edge servers.

\subsection{Lower-Level Coalition Formation Game}
We introduce the hedonic game for coalition formation, propose an auction-aware utility function, and propose the POP algorithm to find a Pareto-optimal partition.

\subsubsection{Hedonic Game}
We formulate the lower-level coalition formation game as a hedonic game where clients use a utility function to evaluate their coalition preferences.

\begin{definition}
A \textbf{hedonic game} $G$ is a pair $(\mathcal{N}, \mathcal{R})$, where $N$ is a finite set of players and $\mathcal{R} = \{\mathcal{R}_i: i \in \mathcal{N}\}$ denotes the preference profile of all the players. Here $\mathcal{R}_i$ denotes the preference profile of player $i$ and is defined by a binary, complete, reflexive, and transitive preference relation $\succeq_i$ on the set of coalitions that player $i$ belongs to, i.e., $\mathcal{N}_i = \{S \subseteq 2^N: i \in S \}$. The strict and indifference preference relations are denoted by $\succ_i$ and $\sim_i$, respectively. 
\end{definition}
\noindent\textbf{Remark}: In our setting, each valid coalition must contain \textbf{one and only one} edge server. 
We also assume \textbf{individual rationality}, meaning each client is at least as well off in a coalition as they would be alone.

Clients aim to maximize their personal utility when choosing a coalition. The utility function of client $i$ in coalition $S$ is:
\begin{equation}\label{eq:client_utility}
    U_i(S) = R_i(S) - C_i(S, \theta_i),
\end{equation}
where $R_i(S)$ denotes the payoff from the coalition and $C_i(S, \theta_i)$ is the training cost as defined in Equation (\ref{eq:client_cost}).

\subsubsection{Auction-Aware Preference Profile Generation}
Unlike existing work on the design of the utility function for single-level games \cite{10054499}, we propose a novel auction-aware utility function, incorporating both lower-level game payoffs and upper-level auction competitiveness.
Based on this utility, we introduce an auction-aware preference profile generation algorithm.
Specifically, we first calculate the auction-aware payoff $R_i(S)$, which can be calculated by
\begin{equation}
    R_i(S) = w_i(S) \cdot \mathbb{E}(R^{(t)}(S)),
\end{equation}
where $\mathbb{E}(R^{(t)}(S))$ is the expected total payoff for coalition $S$ and $w_i(S) = \frac{d_i}{\sum_{j \in S} d_j}$ represents the data contribution proportion.
$\mathbb{E}(R^{(t)}(S))$ is given by 
\begin{equation}
    \mathbb{E}(R^{(t)}(S)) = Pr(S \in \mathcal{M}^{(t)}) \cdot R^{(t)}(S),
\end{equation}
where $Pr(S \in \mathcal{M}^{(t)})$ is the probability of coalition $S$ being selected and $R^{(t)}(S)$ denotes the true payoff.

Since other coalitions' bids are unknown \cite{che1993design}, we estimate $\mathbb{E}(R^{(t)}(S))$ using historical earnings by exponential moving average (EMA):
\begin{equation}\label{eq:est_payoff}
    \hat{R}^{(t)}(S) = 
    \begin{cases}
    \alpha_{e} \hat{R}^{(t-1)}(S)\!+\!(1\!-\!\alpha_{e})R^{(t)}(S), \!\!\!\!& \text{if } S\!\in \!\mathcal{M}^{(t)},\\
    \hat{R}^{(t-1)}(S),              & \text{otherwise},
    \end{cases}
\end{equation}
where $\alpha_{e}$ is the EMA coefficient.
Then, the auction-aware payoff $R_i(S)$ and auction-aware utility $U_i(S)$ are
\begin{equation}\label{eq:client_payoff}
    R_i(S) = \frac{d_i \hat{R}^{(t)}(S)}{\sum_{j \in S} d_j}
\end{equation}
and
\begin{equation}\label{eq:client_utility_exact}
    U_i(S) = \frac{d_i \hat{R}^{(t)}(S)}{\sum_{j \in S} d_j} - C_i(S, \theta_i),
\end{equation}
respectively. The preference function $P_i(S)$, based on utility $U_i(S)$, is
\begin{equation}\label{eq:preference_func}
    P_i(S) = 
    \begin{cases}
    U_i(S), & \text{if } S \notin h(i),\\
    0,              & \text{otherwise},
\end{cases}
\end{equation}
where $h(i)$ contains the coalitions that client $i$ has joined when the client requests to join a new partition. 
This function reduces candidate coalitions, lowering computational complexity \cite{5674046}.
Clients generate preference profiles $\mathcal{R}_i$ by comparing coalitions using $P_i(S)$, resulting in:
$S \succeq_i T \Leftrightarrow P_i(S) \geq P_i(T).$
To reduce the computation complexity and without loss of generality, clients in DualGFL must use $\succ_i$ and $\sim_i$ only in their preference profiles.

\subsubsection{Pareto-Optimal Partitioning Algorithm}
The lower-level hedonic game aims to find a Pareto-optimal partition $\pi^*$ of disjoint coalitions. We propose the POP algorithm to achieve the goal.
We define a partition as follows.
\begin{definition}
    A \textbf{partition} $\pi = \{S_1, S_2, ..., S_K\}$ of $\mathcal{N}$ consists of $K$ disjoint coalitions, where $S_k \neq \emptyset$, $\bigcup_{k=1}^K |S_k| = N$, and $S_k \cap S_l = \emptyset$ for $k \neq l$. Let $\pi(i) = \{S \in \pi: i \in S\}$ be the coalition player $i$ is in. Let $\prod(\mathcal{N})$ denote the collection of all coalition structures in $\mathcal{N}$.
\end{definition}

Pareto dominance is defined as follows.
\begin{definition}
    A partition $\pi$ \textbf{Pareto dominates} $\pi^{\prime}$ if $\pi \succeq_i \pi^{\prime}$ for any player $i$ and if $\pi \succ_j \pi^{\prime}$ for at least one player $j$. A partition $\pi$ is \textbf{Pareto-optimal} if no partition $\pi^{\prime}$ Pareto dominates it.
\end{definition}

\begin{definition}
    Given a preference profile $\mathcal{R}$, a partition $\pi$ is \textbf{perfect} if for any player $i$, $\pi(i)$ is the most preferred coalition, i.e., $\pi(i) \succeq_i S$ for any $S \in \mathcal{N}_i$.
\end{definition}

Perfect partitions, the best for all players, usually do not exist in federated learning. 
However, Pareto optimality can be obtained by finding the perfect partition on relaxed preference profile \cite{AZIZ2013562}.

\vspace{-0.3cm}
\SetKw{define}{define}
\begin{algorithm}[t]
\caption{Pareto-Optimal Partitioning (POP)}\label{alg:hedonic_game}
\KwIn{Preference profiles $\mathcal{R} = \{\mathcal{R}_i: i \in \mathcal{N}\}$}
\KwOut{Pareto-optimal partition $\pi^*$}

Initialize $\mathcal{R}^{\top} = \mathcal{R}$\;
Initialize $\mathcal{R}^{\bot}$ by replacing all $\succ$ in $\mathcal{R}$ with $\sim$\;

$\pi^* = \text{PerfectPartition}(\mathcal{N}, \mathcal{R}^{\bot})$;

\For{$i \in \mathcal{N}$}{
\While{$\mathcal{R}^{\bot}_i \neq \mathcal{R}^{\top}_i$}{    
    $\mathcal{R}_i^{\prime} = \text{Refine}(\mathcal{R}^{\bot}_i, \mathcal{R}^{\top}_i)$;
    
    $\mathcal{R}^{\prime} = (\mathcal{R}^{\bot}_1, ..., \mathcal{R}^{\bot}_{i-1}, \mathcal{R}_i^{\prime}, \mathcal{R}^{\bot}_{i+1}, ..., \mathcal{R}^{\bot}_n)$;

    $\pi = \text{PerfectPartition}(\mathcal{N}, \mathcal{R}^{\prime})$;
        
    \eIf{$\pi \neq \emptyset$}{
        $\pi^* = \pi$;

        $\mathcal{R}^{\bot}_i = \mathcal{R}_i^{\prime}$;
    }{
    $\mathcal{R}^{\top}_i = \mathcal{R}_i^{\bot}$;
    }
}
}

\end{algorithm}

\LinesNumberedHidden
\NoCaptionOfAlgo
\SetKw{define}{define}
\begin{algorithm}
\caption{Refine and PerfectPartition Functions}
\SetKwFunction{FRefine}{Refine}
\SetKwProg{Fn}{Function}{:}{}
\Fn{\FRefine{$\mathcal{R}^{\bot}_i, \mathcal{R}^{\top}_i$}}{
    Replace one $\sim$ in $\mathcal{R}^{\bot}_i$ with $\succ$ towards $\mathcal{R}^{\top}_i$\;
}

\SetKwFunction{FPerfectPartition}{PerfectPartition}
\Fn{\FPerfectPartition{$\mathcal{N}, \mathcal{R}$}}{
    Initialize $\pi$ with empty lists for each edge server\;
    Shuffle edge server preferences and search order\;
    \For{each edge server}{
        \For{each client in server's preference list}{
            \If{server is client's top choice}{
                Add client to server's coalition\;
            }
        }
    }
    \eIf{all clients are allocated}{
        \Return $\pi$
    }{
        \Return $\emptyset$
    }
}

\end{algorithm}
\RestoreCaptionOfAlgo
\LinesNumbered
\vspace{-0.3cm}

We first present the following theorem that characterizes the relationship between the perfect partition and the Pareto-optimal partition. The detailed proof of Theorem \ref{thm:pareto} can be found in \cite{AZIZ2013562}.
\begin{theorem}\label{thm:pareto}
(\textbf{Pareto-Optimal Partition}) Let $(\mathcal{N}, \mathcal{R}^{\top})$ and $(\mathcal{N}, \mathcal{R}^{\bot})$ represent hedonic games where $\mathcal{R}^{\bot} \leq \mathcal{R}^{\top}$, which means that $\mathcal{R}^{\top}$ has some preferences that are more strict than the ones in  $\mathcal{R}^{\bot}$. Suppose $\pi$ is a perfect partition for $\mathcal{R}^{\bot}$. Then $\pi$ is Pareto-optimal for $\mathcal{R}^{\top}$ if and only if there exists a preference profile $\mathcal{R} \in [\mathcal{R}^{\bot}, \mathcal{R}^{\top}]$ such that 
\begin{enumerate}
    \item $\pi$ is perfect for $\mathcal{R}$, and
    \item no partition is perfect for any $\mathcal{R}'$ where $\mathcal{R} < \mathcal{R}' \leq \mathcal{R}^{\top}$.
\end{enumerate}
\end{theorem}

Based on Theorem \ref{thm:pareto} and preference refinement in \cite{AZIZ2013562}, we propose the Pareto-optimal partitioning (POP) algorithm that accepts clients' preference profiles and assigns clients to edge servers to form a Pareto-optimal partition, as shown in Algorithm \ref{alg:hedonic_game}.

Initially, POP sets $\mathcal{R}^{\top}$ to be the original preference profiles and $\mathcal{R}^{\bot}$ to be the relaxed version where all $\succ$ are replaced with $\sim$. The initial perfect partition $\pi^*$ is computed using the \textit{PerfectPartition} function on $\mathcal{R}^{\bot}$, which is guaranteed to exist. 
The \textit{PerfectPartition} function matches the mutual preferences of edge servers and clients by shuffling preferences and iterating through preferred clients, adding them to coalitions if the edge server is the client's top choice.

POP then iteratively refines each player's preference profile from $\mathcal{R}^{\bot}$ towards $\mathcal{R}^{\top}$. The \textit{Refine} function gradually changes indifferences to strict preferences, and \textit{PerfectPartition} seeks a perfect partition for the current profile. If found, the partition is updated. This iterative process ensures the final partition $\pi^*$ is Pareto-optimal for $\mathcal{R}^{\top}$.

\textbf{Discussion of POP: POP returns a Pareto-optimal partition.} Since initial $\pi^*$ always exists and there is no partition perfect for $\mathcal{R}'$ where $\mathcal{R} < \mathcal{R}' \leq \mathcal{R}^{\top}$ when POP terminates. Theorem \ref{thm:pareto} ensures that POP returns a Pareto-optimal partition.
All the Pareto-optimal partitions can be found by changing the iteration order in line 4 of Algorithm \ref{alg:hedonic_game}. For computational efficiency, we only use the first one found.

\textbf{POP algorithm runs in polynomial time.} The computational complexity of POP is dominated by its iterative process and the \textit{PerfectPartition} function. With $N$ clients and $K$ edge servers, the \textit{PerfectPartition} function operates with $O(KN)$ complexity due to nested loops. The outer loop runs $N$ times, and each iteration of the while loop could run up to $(N-1)$ times, leading to an overall complexity of $O(KN^3)$.

\textbf{POP algorithm is strategyproof for general hedonic games with no unacceptability}. This means that clients cannot gain any advantage by misrepresenting their preferences. Since clients are required to rank all possible coalitions without the option to declare any coalition as unacceptable, clients must express a complete and truthful preference order. 
Attempting to manipulate preferences would not yield a more favorable outcome for the player, as the coalition formation process still adheres to the overall preference order. 

\subsection{Upper-Level Multi-Attribute Auction with Resource Constraints}
We introduce the multi-attribute auction model, derive equilibrium bids to maximize coalition utility, and propose an algorithm to solve the coalition selection problem.

\subsubsection{Multi-Attribute Auction Model}
We formulate the competition among coalitions for training participation as a procurement auction, in which the central server is the buyer and $K$ coalitions are the suppliers. The auction proceeds as follows:
1. Each supplier independently places a multi-dimensional bid containing price, qualities, and resource requests.
2. The buyer calculates a score for each bid and selects $M$ winners based on the scores under the resource budget.
3. Winning bids are formalized into contracts, and winners must deliver according to their bids.

Specifically, the bid of coalition $k$ is formulated as 
$B_k = (P_k, \mathbf{Q}_k, E_k),$
where $P_k \in \mathbb{R_+}$ denotes the price, $\mathbf{Q}_k = (q_1^k, q_2^k, ..., q_l^k)  \in \mathbb{R}^l_+$ denotes the quality vector where each entry represents a nonmonetary attribute, such as data volume, data quality, and delivery time. 
$E_k$ denotes the resource request and we use communication bandwidth as the resource in this paper. 

The score for each bid is calculated using a scoring rule $h: \mathbb{R}^{l+1} \to \mathbb{R}$, designed to reflect the buyer's preferences. We use a quasi-linear scoring rule:
\begin{equation}\label{eq:score}
    h(P_k, \mathbf{Q}_k) = \mathbf{Q}_k^T \bm{\alpha} -  P_k,
\end{equation}
where $\bm{\alpha} = (\alpha_1, \alpha_2, ..., \alpha_l) \in \mathbb{R}_+^l$ are quality weights.
The buyer aims to select winners maximizing the total score, while suppliers aim to maximize profits by optimal bids.

\subsubsection{Equilibrium Bids of Coalitions}
In the auction, each supplier's profit is the difference between the price and the cost if its bid wins. The utility of a supplier is
\begin{equation}\label{eq:supplier_utility}
    \pi_k(P_k, \mathbf{Q}_k) =
    \begin{cases}
        P_k - C_k(\mathbf{Q}_k, \theta_k), & \text{if } x_k = 1,\\
        0,              & \text{if } x_k = 0,
    \end{cases}
\end{equation}
where $C_k(\mathbf{Q}_k, \theta_k)$ denotes the total cost to provide qualities $\mathbf{Q}_k$ with private cost factor $\theta_k$ and $x_k \in \{0, 1\}$ is the winning indicator. 
The total cost $C_k(\mathbf{Q}_k, \theta_k)$ is the sum of members' costs:
$C_k(\mathbf{Q}_k, \theta_k) = \sum_{i \in S} C_i(S, \theta_i).$

Suppliers aim to maximize profit by optimizing bids.
Consider that supplier $S$ is one of the winners in the auction with a score $\psi_k$ to fulfill under the scoring rule $h(\cdot)$.
We can formulate the utility maximization problem for the supplier as
\begin{align}
\label{p1}
\tag{P1}
\max_{P_k, \mathbf{Q}_k}\quad & \pi_k(P_k, \mathbf{Q}_k) \\ \tag{\ref{p1}a}
\label{p1a}
\text{s.t.} \quad& h(P_k, \mathbf{Q}_k) = \psi_k.
\end{align}
This can be converted to an unconstrained problem:
\begin{equation}
\label{P2}
\tag{P2}
    \max_{\mathbf{Q}_k}\quad \mathbf{Q}_k^T \bm{\alpha} -  C_k(\mathbf{Q}_k, \theta_k) - \psi_k.
\end{equation}

Based on the scoring auction theory \cite{che1993design}, we give the solution to Problem \ref{P2} in Theorem \ref{thm:eq_bid}.
\begin{theorem}\label{thm:eq_bid}
    (\textbf{Equilibrium Bid}) The multi-attribute auction with a quasi-linear scoring rule has a unique and symmetric equilibrium bid for each supplier, given by
    \begin{align*}
        \mathbf{Q}_k^*(\theta_k) & = \argmax \left(\mathbf{Q}_k^T \bm{\alpha} -  C_k(\mathbf{Q}_k, \theta_k)\right), \\
        P_k^*(\theta_k) & = C_k(\mathbf{Q}_k^*, \theta_k) + \Tilde{P}(\mathbf{Q}_k^*, \theta_k). \numberthis
    \end{align*}
    Here, $\Tilde{P}$ denotes the profit calculated by
    \begin{equation*}
        \Tilde{P}(\mathbf{Q}_k^*, \theta_k) = \int_{\underline{\theta}}^{\overline{\theta}} C_{\theta}(\mathbf{Q}_k^*(t), t) \left[ \frac{1-F(t)}{1-F(\theta_k)} \right]^{N-1} dt, 
    \end{equation*}
    where $\theta_k$ is independently and identically distributed over $[\underline{\theta}, \overline{\theta}]$ and $F(\cdot)$ is the cumulative distribution function.
\end{theorem}

\begin{table*}[h]
\centering
\caption{Performance comparison between DualGFL and baselines. The best performance in each data configuration is in \textbf{bold}. ``Total Score'' does not apply to FedAvg and FedAvgAuc. Except for the metric ``Total Score'', FedAvg is used as the benchmark. For the metric ``Total Score'', FedAvgHed is the benchmark. The improvement ratio in parentheses, such as (38.99x), denotes the improvement with respect to the benchmark. Our DualGFL significantly improves server utility and client utility.}
\begin{adjustbox}{width=0.7\textwidth, totalheight=\textheight, keepaspectratio}
\begin{tabular}{l|l| c c c c c}
\toprule
Dataset & Method & \makecell{Total \\ Score} & \makecell{Avg. Client \\ Quality} & \makecell{Avg. Coalition \\ Quality} & \makecell{Avg. Client \\ Utility} & \makecell{Test \\ Accuracy} \\ 
\midrule \midrule

\multirow{6}{*}{\makecell{FMNIST \\ (0.6)}} 

 & FedAvg  &  - & 1198.19 \footnotesize (1x) & 1198.19 \footnotesize (1x) & 125.92 \footnotesize (1x) & 91.14 \%  \\
 
& FedAvgAuc  &  - &  \textbf{1581.94 \footnotesize (1.32x)} &  1581.94 \footnotesize (1.32x) &  206.59 \footnotesize (1.64x) & 89.54\%  \\ 

& FedAvgHed  & 13.97 \footnotesize (1x) & 1179.91 \footnotesize (0.98x) & 6702.52 \footnotesize (5.59x) & 225.01 \footnotesize (1.79x) & 91.15\%  \\ \cline{2-7}

 & DualGFLStat & 343.84 \footnotesize (24.60x) & 1209.09 \footnotesize (1.01x) & 7833.43 \footnotesize (6.54x) & 272.71 \footnotesize (2.17x) & 90.93\%  \\ 
 
 & DualGFL & \textbf{544.87 \footnotesize (38.99x)*} & 1271.56 \footnotesize (1.06x) & \textbf{12470.54 \footnotesize (10.41x)} & \textbf{483.06 \footnotesize (3.84x)} & \textbf{91.36\%} \\ \midrule \midrule

 \multirow{6}{*}{\makecell{FMNIST \\ (0.1)}} 

 & FedAvg  &  - & 1201.88 \footnotesize (1x) &  1201.88 \footnotesize (1x) & 109.77 \footnotesize (1x) & 88.26\%  \\  
 
& FedAvgAuc  &  - & \textbf{2208.00 \footnotesize (1.84x)} & 2208.00 \footnotesize (1.84x) &  255.21 \footnotesize (2.33x) & 88.08\%  \\ 

& FedAvgHed  & 10.21 \footnotesize (1x) & 1200.10 \footnotesize (1.00x) & 6814.46 \footnotesize (5.67x) & 217.62 \footnotesize (1.98x) & 86.46\%  \\ \cline{2-7} 

 & DualGFLStat & 331.71 \footnotesize (32.50x) & 1225.99 \footnotesize (1.02x) & 7917.18 \footnotesize (6.59x) & 261.28 \footnotesize (2.38x) & 88.12\%  \\ 
 
 & DualGFL & \textbf{516.29 \footnotesize (50.58x)} & 1331.40 \footnotesize (1.11x) & \textbf{12473.63 \footnotesize (10.38x)} & \textbf{447.59 \footnotesize (4.08x)} & \textbf{88.43\%} \\ \midrule \midrule
 
\multirow{6}{*}{\makecell{EMNIST \\ (0.1)}} 

 & FedAvg  &  - & 697.08 \footnotesize (1x) & 697.08 \footnotesize (1x) & 4.43 \footnotesize (1x) & 82.86\%  \\ 
 
& FedAvgAuc  &  - & \textbf{1392.56 \footnotesize (2.00x)} & 1392.56 \footnotesize (2.00x) & 16.42 \footnotesize (3.71x) & 74.92\%  \\ 

& FedAvgHed  & 325.95 \footnotesize (1x) & 698.88 \footnotesize (1.00x) & 6968.68 \footnotesize (10.00x) & 28.01 \footnotesize (6.33x) & 83.22\%  \\ \cline{2-7} 

 & DualGFLStat & 615.96 \footnotesize (1.89x) & 702.11 \footnotesize (1.01x) & 8062.36 \footnotesize (11.57x) & 33.36 \footnotesize (7.54x) & 82.67\%  \\ 
 
 & DualGFL & \textbf{938.72 \footnotesize (2.88x)} & 820.74 \footnotesize (1.18x) & \textbf{12293.74 \footnotesize (17.64x)} & \textbf{55.38 \footnotesize (12.51x)} & \textbf{83.72\%} \\ \midrule \midrule

\multirow{6}{*}{\makecell{CIFAR10 \\ (0.1)}} 

 & FedAvg  &  - & 995.67 \footnotesize (1x) & 995.67 \footnotesize (1x) & 77.14 \footnotesize (1x) & 73.50\%  \\ 
 
& FedAvgAuc  &  - & \textbf{1857.13 \footnotesize (1.87x)} & 1857.13 \footnotesize (1.87x) & 188.29 \footnotesize (2.44x) & 73.93\%  \\ 

& FedAvgHed  & 238.49 \footnotesize (1x) & 983.21 \footnotesize (0.99x) & 5513.51 \footnotesize (5.54x) & 180.18 \footnotesize (2.34x) & 66.30\%  \\ \cline{2-7} 

 & DualGFLStat & 293.33 \footnotesize (1.23x) & 1044.23 \footnotesize (1.05x) & 6761.98 \footnotesize (6.79x) & 230.57 \footnotesize (2.99x) & 73.17\%  \\ 
 
 & DualGFL & \textbf{488.18 \footnotesize (2.05x)} & 1161.65 \footnotesize (1.17x) & \textbf{11240.82 \footnotesize (11.29x)} & \textbf{420.56 \footnotesize (5.45x)} & \textbf{75.17\%} \\
 
 \bottomrule
 
\end{tabular}
\end{adjustbox}
\vspace{-0.35cm}
\label{tab:performance}
\end{table*}

The optimal qualities $\mathbf{Q}_k^*(\theta_k)$ are computed based on the scoring rule, independent of the fulfillment score $\psi_k$. The optimal price includes cost and additional profit based on optimal qualities and cost factor distribution.

\subsubsection{Score Maximization Problem}

The buyer aims to select winners who can maximize the total score, that is:
\begin{align*}
\label{p3}
\tag{P3}
\max_{\{x_k\}}\quad & \sum_{k=1}^K x_k \left(\mathbf{Q}_k^T \bm{\alpha} -  P_k\right) \\
\text{s.t.} \quad& \sum_{k=1}^K x_k  = M, \quad \sum_{k=1}^K x_k E_k \leq E_{max},\\
& x_k = \{0, 1\}, \forall k \in [1, K],
\end{align*}
where $M$ denotes the number of winners and $E_{max}$ denotes the bandwidth budget.

Problem \ref{p3} is a 0-1 Knapsack problem with a cardinality and a resource constraint, which is NP-hard. We propose a greedy algorithm with a score-to-resource ratio to solve the problem.
The algorithm involves two steps:
1. Calculate the score-to-resource ratio for each supplier: $\frac{\mathbf{Q}_k^T \bm{\alpha} -  P_k}{E_k}$, and sort suppliers in descending order.
2. Select winners from the sorted list until $M$ suppliers are chosen or $E_{max}$ is reached.
The algorithm's computational complexity is dominated by the sorting step, resulting in efficient $O(K\log K)$ complexity, making it suitable for large-scale applications.

\section{Experiments}

\textbf{Datasets and Predictive Model:}
We use the FMNIST \cite{xiao2017fashion}, EMNIST \cite{cohen2017emnist}, and CIFAR10 \cite{krizhevsky2009learning} datasets for image classification tasks. 
We implement a shallow convolution neural network with two convolution layers as the classification model in \cite{NEURIPS2023_3baf4eef}.

\textbf{Data Heterogeneity:}
To simulate real-world data in HFL, which is none independently and identically distributed (non-I.I.D.), we use Dirichlet data partitioning  \cite{NEURIPS2023_3baf4eef} to partition original datasets into clients' private datasets. 
We set multiple data configurations: FMNIST (0.1), FMNIST (0.6), EMNIST (0.1), and CIFAR10 (0.1), where values in parenthesis denote the Dirichlet parameters. 

\textbf{Network Simulations:}
We randomly generate graphs to simulate the network topology. Edge servers are placed in a grid with 100 km intervals, and clients are randomly placed within this grid. The maximum coalition size is constrained by a hyperparameter $|S|_{max}$.

\textbf{Metrics:} We evaluate DualGFL using:
\textbf{Test Accuracy}: Prediction accuracy on test data.
\textbf{Total Score}: Total score of winning coalitions. 
\textbf{Average Coalition Quality}: Average quality of winning coalitions. 
\textbf{Average Client Quality}: Average quality of winning clients. 
\textbf{Average Client Utility}: Average utility clients gain from participation.
Higher values in each metric represent better performance. Results are averaged over three runs with different random seeds.

\textbf{Baselines:} We compare our DualGFL with the following baselines:
\textbf{FedAvg} \cite{mcmahan2017communication}: The number of selected clients is adjusted to approximate the number of winning clients in DualGFL.
\textbf{FedAvg Auction (FedAvgAuc)} \cite{9372882}: Only the auction game is applied to FedAvg, where clients place bids directly without forming coalitions.
\textbf{FedAvg Hedonic (FedAvgHed)} \cite{10054499}: Only the hedonic game is applied to FedAvg, where clients form coalitions, and the central server randomly selects coalitions.
\textbf{DualGFL Statistics (DualGFLStat)}: A variant of DualGFL where winning coalitions are randomly selected according to their normalized scores. 

\textbf{System Parameters:}
For FMNIST (0.6), FMNIST (0.1), and CIFAR10 (0.1), we generate $N=50$ clients and $K=9$ edge servers, selecting $M=3$ coalitions in each round. The maximum coalition size is $|S|_{max}=10$. For EMNIST (0.1), we generate $N=1000$ clients and $K=100$ edge servers, selecting $M=5$ coalitions in each round. The maximum coalition size is $|S|_{max}=15$.
Each experiment is conducted in $T=250$ rounds, and clients update the model for $I=3$ epochs using the SGD optimizer with a learning rate of $0.01$ and momentum of $0.9$. The batch size is set to $32$.
The central server uses data size as the quality metric.

\vspace{-0.15cm}
\subsection{Experiment Results}

\textbf{DualGFL shows superior performance in server utility, including total score, average client quality, and average coalition quality.}  As shown in Table \ref{tab:performance}, 
DualGFL achieves improvements of at least 2.05 times in total score, 1.06 times in average client quality, 
and 10.38 times in average coalition quality. 
DualGFLStat shows improvement over FedAvgHed, indicating that score-based selection is superior to random selection for high-quality participants. 

\textbf{DualGFL significantly outperforms baselines in client utility.}
DualGFL provides the highest average client utility, achieving improvement up to 12.51 times over FedAvg.
This significant improvement demonstrates DualGFL's effectiveness in enhancing client welfare. DualGFLStat achieves the second-best performance, suggesting that clients benefit more from participating in the dual-level game of DualGFL than single-level game methods.

\textbf{DualGFL outperforms the baselines in test accuracy across all settings}. Although accuracy improvement is not the primary objective, DualGFL shows strong accuracy performance, especially in the CIFAR10 (0.1) setting with high data heterogeneity, achieving approximately a 1.7\% improvement in accuracy due to the selection of higher-quality coalitions and clients.

\begin{figure} [t]
    \centering
    \includegraphics[width=0.99\linewidth]{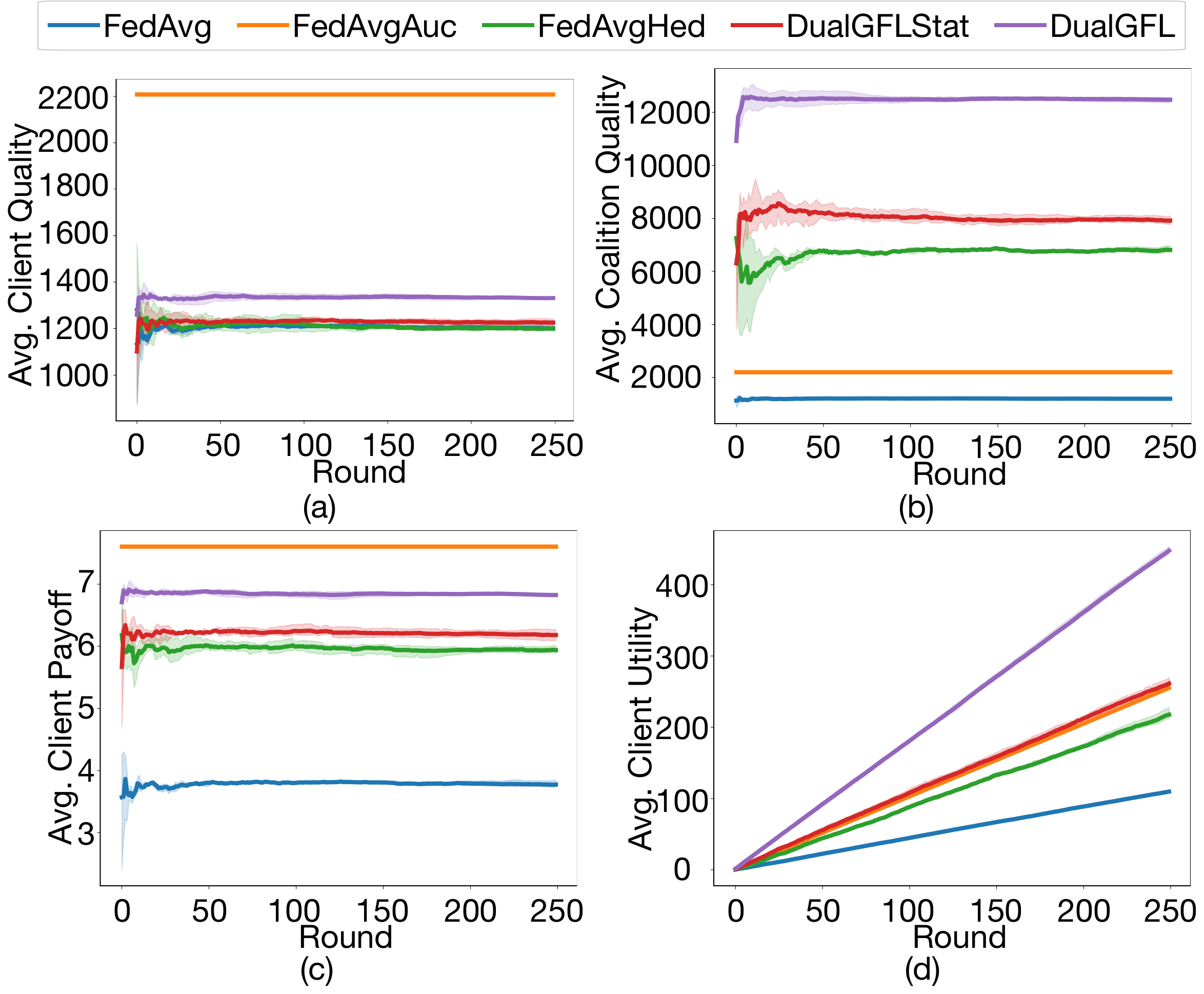}
    \captionsetup{justification=justified, singlelinecheck=false}
    \vspace{-0.25cm}
    \caption{Training dynamics of key metrics. (a), (b), (c), and (d) show the cumulative average of client quality, coalition quality, client payoff, and client utility, respectively.}
    \vspace{-0.5cm}
    \label{fig:fmnist_merge}
\end{figure}

\textbf{Evaluating Training Dynamics} DualGFL significantly improves server and client utility, as shown by key metrics obtained at the end of the training in Table \ref{tab:performance}.
To understand the performance improvement, we present training dynamics for key metrics in the FMNIST (0.1) setting in Figure \ref{fig:fmnist_merge}.
Specifically, Figures \ref{fig:fmnist_merge}(a), \ref{fig:fmnist_merge}(b), \ref{fig:fmnist_merge}(c), and \ref{fig:fmnist_merge}(d) show the cumulative average of client quality, coalition quality, client payoff, and client utility during the training, respectively. 
Shaded areas indicate variations across three runs.

\textbf{DualGFL effectively enhances participant quality.}
There are distinct tiers in average client quality and coalition quality from Figure \ref{fig:fmnist_merge}(a) and \ref{fig:fmnist_merge}(b), with FedAvgAuc and DualGFL consistently selecting better quality clients. Despite having the highest client quality, FedAvgAuc shows low coalition quality due to the lack of coalition formation.

\textbf{Clients gain higher payoffs by joining higher-quality coalitions.}
The average client payoff stabilizes as training progresses, as shown in Figure \ref{fig:fmnist_merge}(c), with consistent differences among methods, which explains the stable and linear increase in client utility as shown in Figure \ref{fig:fmnist_merge}(d). The distribution of client payoff mirrors coalition quality except for FedAvgAuc, indicating clients benefit from participating in high-quality coalitions.

\begin{figure}[t]
    \centering
    \includegraphics[width=0.99\linewidth]{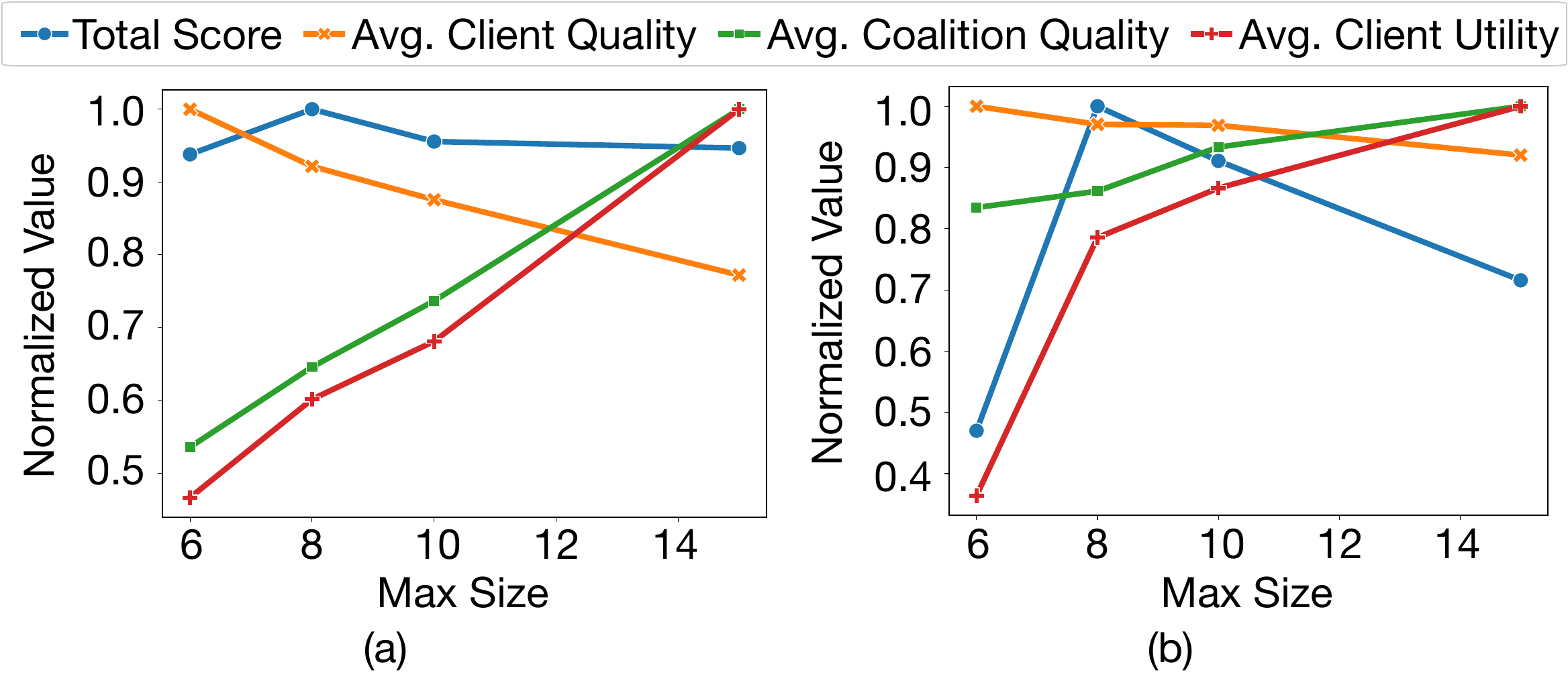}
    \captionsetup{justification=justified, singlelinecheck=false}
    \vspace{-0.25cm}
    \caption{Impact of max size $|S|_{max}$ on the performance of the coalition selection. (a) and (b) show the performance of DualGFL and DualGFLStat, respectively.}
    \vspace{-0.5cm}
    \label{fig:ab_max_size_fmnist}
\end{figure}
\textbf{Ablation Study:}
Increasing the maximum coalition size $|S|_{max}$ affects the performance of DualGFL. We conduct an ablation study on $|S|_{max}$ in FMNIST (0.1) setting. Smaller values 
produce more uniformly sized coalitions. 
We tested $|S|_{max} \in [6, 8, 10, 15]$.
Figure \ref{fig:ab_max_size_fmnist} shows normalized values for the total score, average client quality, average coalition quality, and average client utility across different $|S|_{max}$.

\textbf{Larger $|S|_{max}$ produces higher average coalition quality but lower client quality.}
Larger $|S|_{max}$ increases size discrepancy among coalitions.
Coalitions with size advantage are prone to win the auction, leading to higher average coalition quality.
However, larger coalitions attract more ``free riders'' with lower quality, causing a decrease in average client quality but still increasing average client utility. The total score peaks at $|S|_{max}=8$, balancing client movement among coalitions and avoiding the Matthew effect, where larger coalitions have a dominant advantage.

\section{Conclusion}\label{sec:conclusions}
We introduced DualGFL, the first federated learning framework combining a dual-level game to enhance both server and client utilities. The lower level uses a hedonic game for coalition formation, where we proposed an auction-aware utility function and a Pareto-optimal partitioning algorithm. The upper level formulates a multi-attribute auction game with resource constraints, deriving equilibrium bids and solving a score maximization problem for the central server. 
Experiments show that DualGFL consistently outperforms single-game baseline models across various metrics.

\section{Acknowledgments}
This work was supported in part by the NSF under Grant No. 1943486, No. 2246757, No. 2315612, and No. 2332011, and in part by a grant from BoRSF under contract LEQSF(2024-27)-RD-B-03.

\bibliography{main}

\clearpage
\appendix
\setcounter{figure}{0}    
\setcounter{table}{0}     
\setcounter{algocf}{0}    

\section*{Summary of Key Notation}

The key notation in this work is summarized in Table \ref{table:notation}.

\begin{table}[htbp]
\renewcommand{\thetable}{A-\arabic{table}}
  \centering
  \caption{List of Key Notation.}
  \label{table:notation}
  \begin{adjustbox}{width=0.95\columnwidth, totalheight=\textheight, keepaspectratio}
  \begin{tabular}{l|l}
    \hline
    \textbf{Symbol} & \textbf{Description} \\
    \hline
    $f(x), F_i(x)$ & Global and local loss functions \\
    $f^*$ & Minimum global loss \\
    $d_i$ & Ratio of the local data volume of client $i$ \\
    $\mathbf{p}$ & Client-selection probability \\
    $\mathcal{M}^{(t)}$ & Participants in round $t$ \\
    $N$ & Number of clients\\    
    $K$ & Number of coalitions \\
    $M$ & Number of selected coalitions per round \\
    $\xi_j^i$ & $j$-th data point in the dataset of client $i$ \\
    $T, I$ & Number of global rounds and local epochs \\
    $c_{i, cp}, c_{i, cm}(S)$ & Computation and communication cost of client $i$ \\
    $\theta_i$ & Cost factor of client $i$ \\
    $\alpha_e$ & EMA coefficient of payoff estimate \\
    $\mathcal{R}$ & Preference profile\\
    $\hat{R}^{(t)}(S)$ & Estimated payoff for coalition $S$\\
    $R_i(S), U_i(S)$ & Payoff and utility for client $i$\\
    $\pi^*$ & Pareto-optimal partition\\
    $\bm{\alpha}$ & Quality weights\\
    $B_k, P_k, \mathbf{Q}_k, E_k$ & Bid, price, qualities, and resource for coalition $k$\\
    $\mathbf{Q}_k^*, P_k^*$ & Equilibrium qualities and price for coalition $k$\\
    \hline
  \end{tabular}
  \end{adjustbox}
\end{table}

\section*{Auction-Aware Preference Profile Generation Algorithm}\label{app:preference_profile}

The proposed auction-aware preference profile generation algorithm is shown in Algorithm \ref{alg:preference_profile}.

\begin{algorithm}
\renewcommand{\thealgocf}{A-\arabic{algocf}}
\SetAlgoLined
\caption{Auction-Aware Preference Profile Generation}\label{alg:preference_profile}
\KwIn{Client set $\mathcal{N}$, EMA coefficient $\alpha_e$}
\KwOut{$\mathcal{R} = \{\mathcal{R}_i: i \in \mathcal{N}\}$}

\For{each client $i \in \mathcal{N}$ in parallel}{
\For{$t=1, 2, ..., T$}{
    
    \For{each coalition $S\in \mathcal{N}_i$}{
        Update coalition payoff $\hat{R}^{(t)}(S)$ by (\ref{eq:est_payoff})\;
    
        Compute client payoff $R_i(S)$ by (\ref{eq:client_payoff})\;

        Compute total cost $C_i(S)$ by (\ref{eq:client_cost})\;
        
        Compute client utility $U_i(S)$ by (\ref{eq:client_utility_exact})\;
        
        Update preference  $P_i(S)$ by (\ref{eq:preference_func})\;
    }
    
    Sort the coalitions based on $P_i(S)$ to generate the preference profile $\mathcal{R}_i$\;
}
}
\end{algorithm}

\newpage

\section*{Greedy Algorithm with Score-to-Resource Ratio}\label{app:greedy}

We present the greedy algorithm to solve the score maximization problem in Algorithm \ref{alg:greedy}.

\begin{algorithm}
\renewcommand{\thealgocf}{A-\arabic{algocf}}
\caption{Greedy Algorithm With Score-to-Resource Ratio}\label{alg:greedy}
\KwIn{Bids $\{(P_k, \mathbf{Q}_k, E_k): k \in [1, K]\}$, resource budget $E_{max}$, number of winners $M$}
\KwOut{Set of winners $\mathcal{M}$}

Initialize $\mathcal{M} = \emptyset$, $\Tilde{E}= 0$\;

\For{k = 0, 1, ...K}{
Calculate score-to-resource ratio $SR_k = \frac{\mathbf{Q}_k^T \bm{\alpha} -  P_k}{E_k}$\;
}

Sort suppliers in descending order based on $SR_k$ to get list $\{SR_k\}$\;

\For{each supplier $k$ in $\{SR_k\}$}{
    \If{$\Tilde{E} \leq E_{max}$ \textbf{and} $|\mathcal{M}| < M$}{
        $\mathcal{M} = \mathcal{M} \cup \{k\}$\;
        
        $\Tilde{E} = \Tilde{E} + E_k$\;
    }
}

\end{algorithm}

\section*{Proof of Theorem \ref{thm:eq_bid}}\label{app:proof_eq_bid}

\begin{proof}
To prove Theorem \ref{thm:eq_bid}, we first establish that for each supplier, the optimal quality is selected by $\mathbf{Q}^*(\theta) = \argmax \left(\mathbf{Q}^T \bm{\alpha} -  C(\mathbf{Q}, \theta)\right)$.
We proceed by contradiction. For simplicity, we omit the subscript $k$ in the following notation.

Assume, contrary to the claim, that there exists an equilibrium bid $(P, \mathbf{Q})$ such that $\mathbf{Q} \neq \mathbf{Q}^*$. We will demonstrate that there exists an alternative bid $(P^\prime, \mathbf{Q}^\prime)$, where $\mathbf{Q}^\prime = \mathbf{Q}^*$ and $P^\prime = P + (\mathbf{Q}^\prime - \mathbf{Q})^T \bm{\alpha}$, which strictly dominates the original bid $(P, \mathbf{Q})$. That is, we will show that $\pi(P^\prime, \mathbf{Q}^\prime) > \pi(P, \mathbf{Q})$.

We first show that the quasi-linear scoring rule $h(\cdot)$ assigns the same score to both bids:
\begin{align*}
h(P^\prime, \mathbf{Q}^\prime) 
& = (\mathbf{Q}^\prime)^T \bm{\alpha} -  P^\prime\\
& = (\mathbf{Q}^\prime)^T \bm{\alpha} -  (P + (\mathbf{Q}^\prime - \mathbf{Q})^T \bm{\alpha})\\
& = \mathbf{Q}^T \bm{\alpha} - P \\
& = h(P, \mathbf{Q}).
\end{align*}

Next, we consider the profit $\pi$ associated with these bids:
\begin{align*}
& \pi(P^\prime, \mathbf{Q}^\prime) \\
& = (P^\prime - C(\mathbf{Q^\prime}, \theta)) \text{Prob}(\text{win}|h(P^\prime, \mathbf{Q}^\prime)) \\
& = (P + (\mathbf{Q}^\prime - \mathbf{Q})^T \bm{\alpha} - C(\mathbf{Q}^\prime, \theta)) \text{Prob}(\text{win}|h(P^\prime, \mathbf{Q}^\prime)) \\
& = (P - C(\mathbf{Q}, \theta)) \text{Prob}(\text{win}|h(P, \mathbf{Q})) + \\
& \left( (\mathbf{Q}^\prime\!-\!\mathbf{Q})^T \bm{\alpha}\!-\!C(\mathbf{Q}^\prime, \theta)\!+\! C(\mathbf{Q}, \theta) \right) \text{Prob}(\text{win}|h(P^\prime, \mathbf{Q}^\prime)).
\end{align*}
Given the assumption that
\begin{equation*}
    (\mathbf{Q}^\prime - \mathbf{Q})^T \bm{\alpha} -  C(\mathbf{Q}^\prime, \theta) +  C(\mathbf{Q}, \theta) > 0,
\end{equation*}
it follows that
\begin{equation*}
    \pi(P^\prime, \mathbf{Q}^\prime) > \pi(P, \mathbf{Q}),
\end{equation*}
This inequality contradicts the assumption that $(P, \mathbf{Q})$ is an equilibrium bid. Therefore, in an equilibrium bid, the optimal quality is chosen by $\mathbf{Q}^*(\theta) = \argmax \left(\mathbf{Q}^T \bm{\alpha} -  C(\mathbf{Q}, \theta)\right)$.

Finally, with the optimal quality $\mathbf{Q}^*$ determined, the equilibrium price $P^*$ can be derived using standard results from first-price auctions \cite{Riley1981Optimal}.
\end{proof}

\section*{Datasets}

\textbf{FMNIST}
The FMNIST \cite{xiao2017fashion} dataset consists of a training set of 60,000 examples and a test set of 10,000 examples. Each example is a 28x28 grayscale image associated with a label from 10 classes.
To simulate various non-I.I.D scenarios in federated learning, we use Dirichlet parameters 0.6 and 0.1 to partition FMNIST data into private datasets of 50 clients.
By adjusting the Dirichlet parameter, we can simulate various levels of data heterogeneity, leading to diverse data volumes and class distributions among clients.
Lower values correspond to higher data heterogeneity.

\textbf{EMNIST}
The EMNIST dataset \cite{cohen2017emnist} is a collection of handwritten character digits, providing a set of 28x28 pixel images. ``ByClass'' split of the dataset is used in our work, including a total of 814,255 characters in 62 unbalanced classes. 
Original data are partitioned into 1,000 clients by Dirichlet distribution, each with its own training, validation, and test datasets. 
Collectively, the training datasets contain 671,585 instances, while the validation and test datasets each comprise 77,483 instances.

\textbf{CIFAR10}
The CIFAR10 dataset \cite{krizhevsky2009learning} consists of 60,000 32x32 RGB images in 10 classes, with 6,000 images per class. There are 50,000 training images and 10,000 test images.
The original data are partitioned into 50 clients' private datasets by Dirichlet distribution.

\textbf{Data Preprocessing}
Images are randomly augmented by preprocessing methods, such as cropping and horizontal flipping.

\section*{Computing Resources for the Experiment}

DualGFL is built with Flower \cite{beutel2020flower} framework and network simulation package: NetworkX.
The experiments are run in the simulation environment on high-performance computing clusters with the following hardware:
\begin{itemize}
    \item 24-core Intel CPU
    \item 96 GB of memory
    \item 1 NVIDIA V100S GPU (with 32 GB memory)
    \item 1.5 TB SSD drive
\end{itemize}

\newpage

\section*{Additional Results}
\setcounter{figure}{0}    
\setcounter{table}{0}     
\setcounter{algocf}{0}    

\begin{figure} [h]
\renewcommand{\thefigure}{A-\arabic{figure}}
    \centering
    \includegraphics[width=0.99\linewidth]{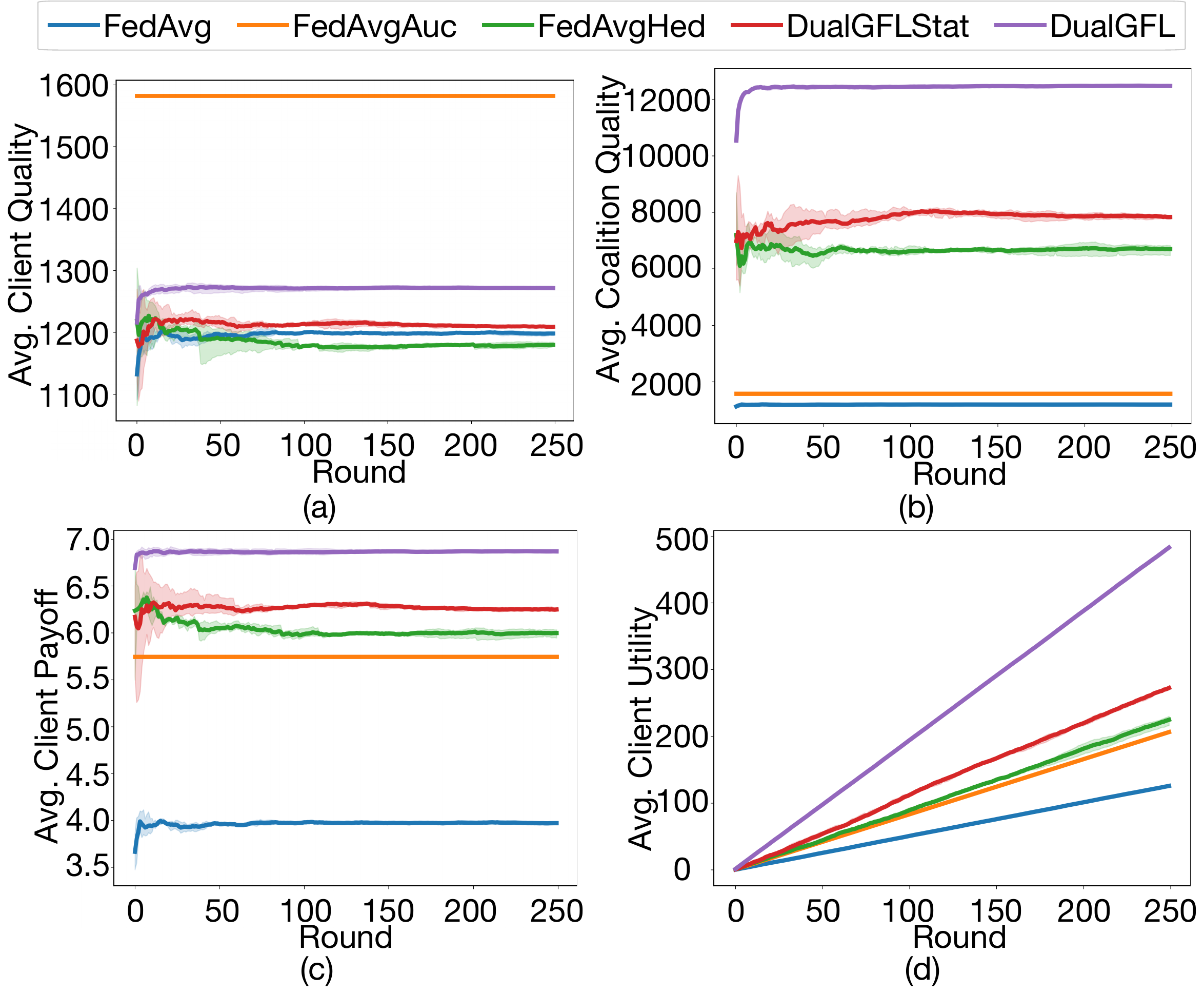}
    \captionsetup{justification=justified, singlelinecheck=false}
    \caption{Training dynamics of key metrics in FMNIST (0.6) setting.}
    \label{fig:fmnist06_merge}
\end{figure}

\begin{figure} [h]
\renewcommand{\thefigure}{A-\arabic{figure}}
    \centering
    \includegraphics[width=0.99\linewidth]{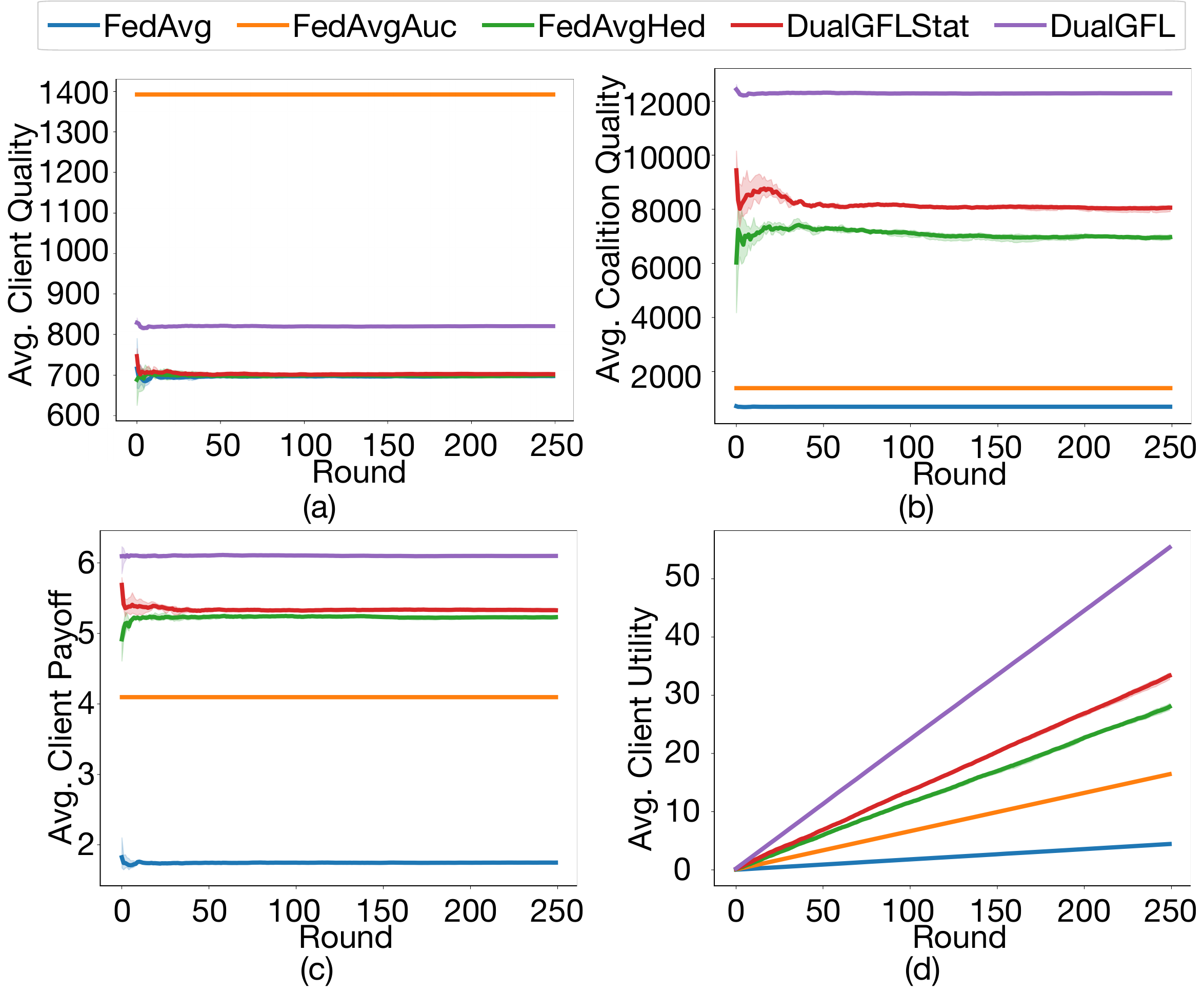}
    \captionsetup{justification=justified, singlelinecheck=false}
    \caption{Training dynamics of key metrics in EMNIST (0.1) setting.}
    \label{fig:emnist01_merge}
\end{figure}

\begin{figure} [h]
\renewcommand{\thefigure}{A-\arabic{figure}}
    \centering
    \includegraphics[width=0.99\linewidth]{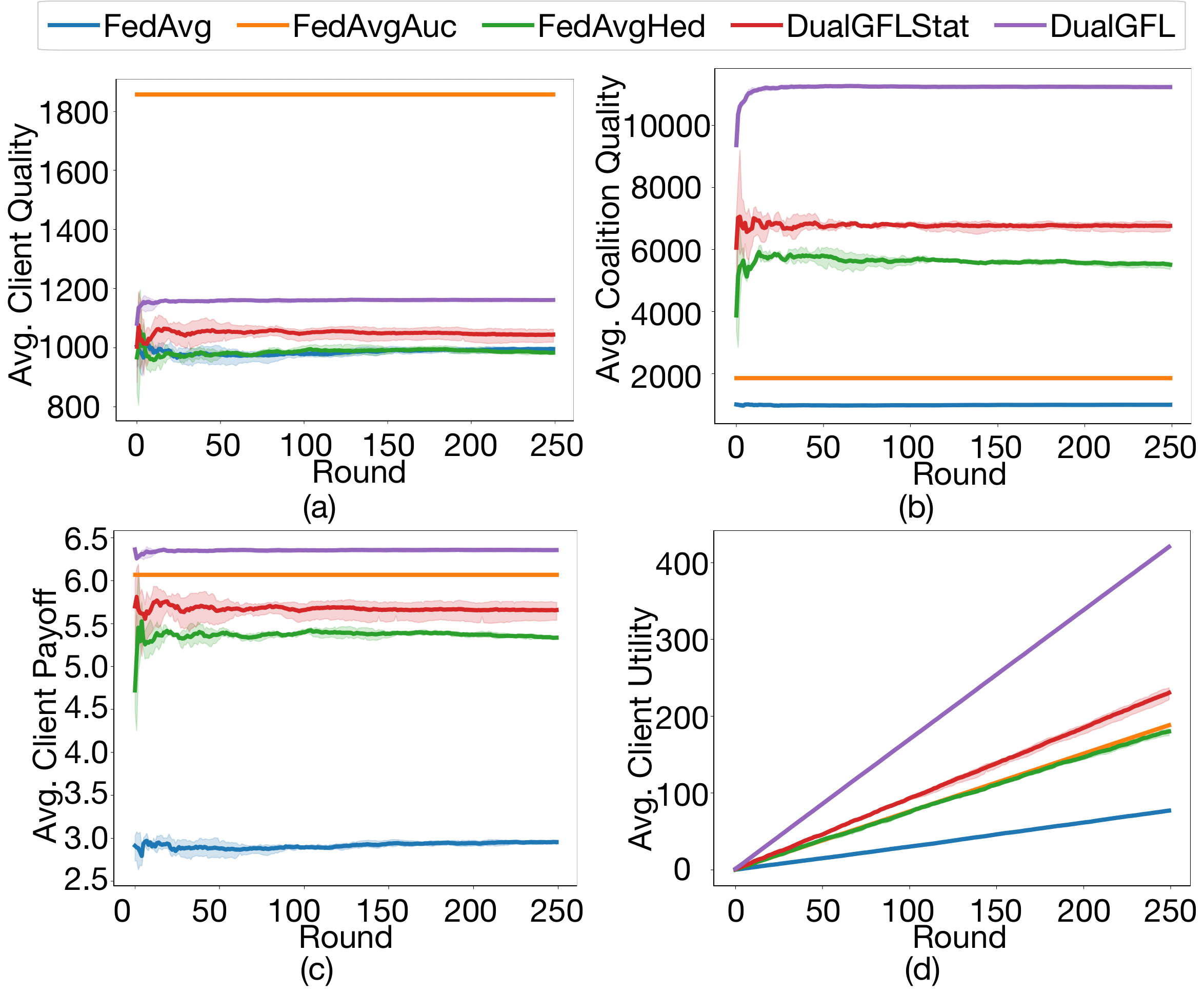}
    \captionsetup{justification=justified, singlelinecheck=false}
    \caption{Training dynamics of key metrics in CIFAR10 (0.1) setting.}
    \label{fig:cifar01_merge}
\end{figure}

We present additional results of training dynamics of DualGFL on data configurations: FMNIST (0.6), EMNIST (0.1), and CIFAR10 (0.1) in Figure \ref{fig:fmnist06_merge}, \ref{fig:emnist01_merge}, and \ref{fig:cifar01_merge}, respectively.
Subfigure (a), (b), (c), and (d) show the cumulative average of client quality, coalition quality, client payoff, and client utility, respectively.

\newpage



\end{document}